\documentclass[11pt,a4paper]{article} 
\usepackage{epsfig} 

\title{
The World of Baby Skyrmions: \\ 
Numerical Studies of (2+1)D Topological Skyrme-like Solitons \\
(Talk given at School of Subnuclear Physics, Erice 1999)
}  
\author{Tom Weidig,
\\Theoretical Physics,\\UMIST,  \\ Manchester M60 1QD 
\\email: tom.weidig@physics.org} 

\newcommand{\ph}{\vec{\phi}} \newcommand{\pha}{\phi_{a}}
\newcommand{\dmu}{\partial_{\mu}} \newcommand{\umu}{\partial^{\mu}}
\newcommand{\dnu}{\partial_{\nu}} \newcommand{\unu}{\partial^{\nu}}

\begin{document}
\maketitle  \abstract{This is one of the `New Talents' seminars at
`Erice International School of  Subnuclear Physics 1999' and looks at
numerical  studies of (2+1)D topological Skyrme-like solitons; the
baby skyrmions. We explain the  concept of integrable and topological
solitons. Then we introduce the non-linear  $\sigma$ model in 2D in
order to discuss the 3D nuclear Skyrme model. We explain that the baby
Skyrme models can be viewed as a (2+1)D version of the  Skyrme
model. We describe the numerical methods needed to study the baby
skyrmions. Finally, we present some results on skyrmion scattering and
our work on skyrmions with higher topological charge.}

\section{Introduction to solitons}

The concept of solitons goes back to J. Scott Russell's discovery of
'great waves of translations' on the Edinburgh-Glasgow canal in
1834. He observed that a boat which
suddenly stops creates a single, localised water wave; now named `a
solitary wave'.  Solitons are those solitary waves which keep their identity
after interactions.   Later work by Boussinesq (1871), by Lord
Rayleigh (1876) and by Korteweg and de Vries (1895) shows that this
solitary wave can be described by the KdV equation:
\begin{equation}
u_t+u_x+uu_x+u_{xxx}=0.
\end{equation} 
The stability of the KdV solitary wave is due to the fine dynamical
balance between  the non-linear and the dispersive term in the
differential equation. In 1968, Muira, Gardner and Kruskal showed that
the KdV equation conserves an infinite number of quantities of
motion. The KdV system is an integrable system.  Techniques like
Lax pair, Inverse Scattering and B\"acklund transformations are available and
they enable us to find explicit  solutions. This makes integrable solitonic
systems good toy models to probe physical ideas; albeit not very
realistic  ones. The main weak points are: finding integrable systems in
other than (1+1)D, imposing Lorentz invariance, adding a small perturbative
term without breaking integrability and ensuring the presence of annihilation 
processes for solitons. (see \cite{soliton:intro} for an introduction to 
solitons)

The simplest example of a topological soliton is the Sine-Gordon
model whose field is described by
\begin{equation}
{\cal L}=\dmu\phi\umu\phi + (1-\cos\phi).
\end{equation}
The lagrangian is invariant under $\phi\longrightarrow\phi+2\pi n, n
\epsilon  {\bf Z}$. $\phi$ is an angle in field space, the circle
$S^1$. Let us re-emphasise our definition of a soliton: it is a
localised finite-energy field configuration which keeps localised
after interactions. The field has to go to the vacuum sufficiently
fast for the soliton to be localised and of finite energy. Therefore,
we can identify the spatial infinity in each direction with one single
point and compactify the one-dimensional space ${\cal R}^1$ to ${\cal
S}^1$. The field theory of the Sine-Gordon model can be described by
the  map
\begin{equation}
 {\cal M}(t):  S^1 \longrightarrow S^1    
\end{equation}
at a given time $t$. This non-trivial mapping gives us the possibility
to partition the space of all possible field configurations into
equivalence classes having the same topological charge or winding
number. We can visualise this concept with a belt. We can  trivially
close it or we can twist one side by 180 degrees and close it or we
can anti-twist it by 180 degrees i.e. twist it by -180 degrees and
close it. We introduce a twist into the belt which cannot be un-done
unless you open the belt. Topological solitons are very much twisted
field configurations fixed by boundary conditions. In a formal
language, we get a non-trivial homotopy group  $\Pi_1(S^1)={\bf Z}$
from the map of the Sine-Gordon field which allows us to add or
subtract field configurations.  For example, if we twist the belt
twice and anti-twist it twice, we get back to an untwisted belt: very
much like an annihilation process in particle physics.  The `twist'
i.e. the topological charge is fixed by boundary conditions and
conserved. Further, we can easily impose Lorentz invariance by only
using covariant lagrangian terms and work in any dimension by
construction.  The weak points of topological solitons are: generally
they are non-integrable systems, few techniques are available (only
Ans\"atze, topological bounds, etc.) and a computational approach is
required. The Sine-Gordon model is a very good toy model, because it is
topological and integrable at the same time: it is close to real
physics and exactly solvable.  Further, we can compare our numerical
methods with exact results.  (see \cite[section 2.5]{raj} for more
details)

\section{Non-linear $\sigma$ model in 2D}

The non-linear $\sigma$ model is described by
\begin{equation}
{\cal L}=\dmu\ph\cdot\umu\ph
\end{equation}
where $\ph$ is a three-dimensional field vector on the sphere $S^2$
i.e. $\ph\cdot\ph=1$. We could also use a complex field $W$ which
represents the stereographic projection of the sphere. Again, we have
a map
\begin{equation}
 {\cal M}(t):  S^2 \longrightarrow S^2 
\end{equation}
which leads to the second homotopy group $\Pi_2(S^2)={\bf Z}$. The
existence of the topological charge i.e. the twisted field
configuration representing a soliton is ensured by topology. However,
we also need to make sure that the soliton is dynamically stable
i.e. that it has a stable scale. Derrick's theorem provides a
necessary but not sufficient condition for stability:
\begin{eqnarray}
\left.\frac{dE[\tilde{\phi}(\lambda x)]}{d\lambda}\right|_{\lambda=1}&=&0 \cr
\left.\frac{d^2E[\tilde{\phi}(\lambda x)]}{d\lambda^2}\right|_{\lambda=1}&\ge&0 .
\end{eqnarray}
The lagrangian is scale invariant and does not satisfy the second
condition. A change of scale does not change the energy. Numerical
simulations are unstable, because numerical errors change the scale of
the soliton without loss of energy. Another important feature is the
existence of a topological bound, the Bogomolnyi bound
\begin{equation}
E\geq4\pi n.
\end{equation}
If we were to extend the model to (3+1)D, the $\sigma$ model term would
expand the soliton: another term needs to be added to ensure
stability. (See  \cite[section 3.2 and 3.3]{raj} for further details)

\section{The nuclear Skyrme model in 3D}

In the early 60's, Skyrme came up with a theory to describe the hadronic
spectrum i.e.\ a theory of mesons and baryons \cite{sky:61}. Its lagrangian 
\begin{equation}
L=\partial_{\mu}\ph\cdot\partial^{\mu}\ph
-\theta_{S}\left[(\partial_{\mu}\vec{\phi} \cdot
\partial^{\mu}\vec{\phi})^{2} -(\partial_{\mu}\vec{\phi} \cdot
\partial_{\nu}\vec{\phi}) (\partial^{\mu}\vec{\phi} \cdot
\partial^{\nu}\vec{\phi})\right] 
\end{equation}
is a straightforward extension of the non-linear $\sigma$ model plus
the addition of a fourth-order term called the Skyrme term. We need to
include this extra term to ensure stability according to Derrick's
theorem. The mapping becomes
\begin{equation}
 {\cal M}(t):  S^3 \longrightarrow S^3.
\end{equation}
 In effect, the most commonly used notation is in terms of a
coordinate set on the $SU(2)$ group manifold.  Skyrme had the
ingenious idea to construct an effective field theory of mesons where
the baryons are the topological solitons of the theory. Baryon
conservation is equivalent to the conservation of the topological
charge. His ideas were put aside by the success of QCD in the
description of hadrons in terms of quarks.  A crucial step forward in
the understanding of QCD is the emergence of the idea of confinement.
In QED, the running coupling constant decreases with distance of
interaction and a perturbation expansion is very effective. In QCD,
the inverse is true: only at small distance ergo high energy does the
coupling become small and the quarks are free particles. At low
energies, people believe that the quarks form colourless
colour-singlet states: the hadrons. This must be true as there is no
experimental evidence to the contrary. So far, no-one has come up with
a decent way of showing confinement in QCD\@. The lack  of a
small parameter to do a perturbation expansion is evident. In the mid
70's, t'Hooft \cite{thooft:2} generalised QCD to be invariant under a
$SU(N_C)$ gauge group. He realised that the quantum treatment of QCD
simplified considerably and it is possible to derive some qualitative
statements in the large $N_C$ and low energy limit.  This comes from
the fact that non-planar diagrams and internal quark loops are
suppressed by a factor of $N_C^{-2}$ respectively $N_C^{-1}$. For
large $N_C$, QCD is a weakly interacting theory of mesons with the
meson-meson interaction of order $N_C^{-1}$. For
$N_C\rightarrow\infty$, QCD is therefore a theory of mesons which are
free and non-interacting. Witten \cite{witten} realised that weakly
coupled theories may exhibit non-perturbative states like solitons or
monopoles. He showed that baryons behave as if they are solitons in a
large-$N_C$ meson theory. The derivation of effective actions from the
QCD lagrangian is an unsolved problem. The Skyrme model is the
simplest of a candidate theory for the low-energy effective lagrangian
of QCD\@. Of course, one may add some higher order correction
terms. See \cite[chapter 9]{mrs} for a simple review on the relation
between QCD and the Skyrme model.

There are difficulties surrounding the quantization of the
(3+1)-dimensional Skyrme model.  First of all, the quantisation is
ambiguous, because the theory is not renormalisable. One can do a
renormalisation to a given order, but it will depend on the
scheme. Therefore, it is essential to compare the scheme-dependent
results with experimental data.  In 1981, Adkins, Nappi and Witten
\cite{anw} partially overcame these problems by using a spinning-top
approximation; the 1-skyrmion i.e. skyrmion of charge one is radially symmetric and rotating
without deformation. Effectively, they only include the rotational
zero mode of  the 1-skyrmion. This procedure allows to quantise the
1-skyrmion as a proton  and gives reasonable agreement
with experiments. The numerical work by Braaten et al. 
and Battye and Sutcliffe \cite{multi:num} on the structure of
classical multi-skyrmions, i.e. skyrmions of charge two and more, supports 
further the idea that an
appropriate quantization around these minimal-energy solutions for a
given topological sector could lead to an effective description of
atomic nuclei.  However, the calculation of quantum properties of the
multi-skyrmions is very difficult, because these minimal-energy
solutions are not radially symmetric, for example. This is rather
frustrating, for the claim that the Skyrme model descending from a
large N-QCD approximation models mesons, baryons and higher nuclei is
a very compelling one. There have been various attempts to extract
quantum properties, notably by Walet \cite{walet} and Leese et al.
\cite{manton}. A recent zero mode quantization of multi-skyrmions from
charge four to nine has been undertaken by Irvine
\cite{irvine}. Reliable quantitative results were not to be expected,
but even the correct quantum number for the ground states of some
nuclei, skyrmions with odd charge 5, 7 and 9, could not be obtained.
All these attempts rely on theoretical approximations like
moduli-space or rational map Ansatz and mostly include only zero modes.

\section{The baby Skyrme models}

The nuclear Skyrme model has two big problems: quantization is
ambiguous and very hard and numerical simulations are computationally
very intense even on supercomputers. It makes sense to look for a
(2+1) version. The baby Skyrme model is a modified version of the
$S^2$ sigma model and the lagrangian is
\begin{equation}
L=\partial_{\mu}\ph\cdot\partial^{\mu}\ph
-\theta_{S}\left[(\partial_{\mu}\vec{\phi} \cdot
\partial^{\mu}\vec{\phi})^{2} -(\partial_{\mu}\vec{\phi} \cdot
\partial_{\nu}\vec{\phi}) (\partial^{\mu}\vec{\phi} \cdot
\partial^{\nu}\vec{\phi})\right] - \theta_V V(\ph)
\end{equation}
The addition of a potential and a Skyrme term to the lagrangian
ensures stable solitonic solutions. The Skyrme term has its origin
from the nuclear Skyrme model and the baby Skyrme model can therefore
be viewed as its (2+1)D analogue. It plays the role of a toy model to
explore Skyrme-like properties. However, in recent years people have
also tried to use the baby Skyrme model to model aspects of the
Quantum Hall Effect.  Further, in (2+1)D, a potential term is
necessary in the baby Skyrme models to ensure stability of skyrmions;
this term is optional in the (3+1)D nuclear Skyrme model. We find the
minimal-energy solution for topological charge one by using the
Hedgehog Ansatz;
\begin{equation}
\vec{\phi}=\left(\matrix{\sin[f(r)]\sin(n\theta-\chi)\cr\sin[f(r)]
\cos(n\theta-\chi)\cr\cos[f(r)]\cr}\right).
\end{equation}
Note that $n$ is a non-zero integer (it is the topological charge),
$\theta$ the polar angle, $\chi$ a phase shift and $f(r)$ the profile
function satisfying certain boundary conditions.  We end up with a
one-dimensional energy functional
\begin{equation}
E=(4\pi)\frac{1}{2}\int_{0}^{\infty} rdr\left( {f^{'}}^{2} +
n^{2}\frac{\sin^{2}f}{r^{2}}(1+2\theta_{S}{f^{'}}^{2}) +
\theta_{V}\mbox{\~{V}}(f)\right).
\end{equation}
The corresponding Euler-Lagrange equation with respect to the
invariant  field $f(r)$ leads to a second-order ODE,
\begin{eqnarray}
  \left(r+\frac{2\theta_{S}n^{2}\sin^{2}f}{r}\right)f^{''} +
  \left(1-\frac{2\theta_{S}n^{2}\sin^{2}f}{r^{2}}+
  \frac{2\theta_{S}n^{2}\sin{f}\cos{f} f{'}}{r} \right)f^{'} \nonumber
  \\ -\frac{n^{2}\sin f\cos f}{r}
  -r\frac{\theta_{V}}{2}\frac{d\mbox{\~{V}}(f)}{df} =0,
\end{eqnarray} 
which we re-write in terms of the second derivative of the profile
function:
\begin{equation}
\label{profile} f^{''}={\cal F}(f,f^{'},r). 
\end{equation}
The profile function $f(r)$ is a static solution of the baby Skyrme
model. These static solutions are certainly critical points, but not
necessarily global minima. However, it is reasonable to assume that
the  hedgehog solution for topological charge one is the
minimal-energy solution.

Taking into account the $S^2$ constraint, the equation of motion is
\begin{center}
\begin{eqnarray}
  \dmu\umu\pha-(\ph\cdot\dmu\umu\phi)\pha -2\theta_{S}[
  (\dnu\ph\cdot\unu\ph) \dmu\umu\pha
  +(\dmu\unu\ph\cdot\umu\ph)\dnu\pha \nonumber \\
  -(\unu\ph\cdot\umu\ph)\dnu\dmu\pha
  -(\dnu\unu\ph\cdot\umu\ph)\dmu\pha
  +(\dmu\ph\cdot\umu\ph)(\dnu\ph\cdot\unu\ph)\pha \nonumber \\
  -(\dnu\ph\cdot\dmu\ph)(\unu\ph\cdot\umu\ph)\pha]
  +2\theta_{V}\frac{dV}{d\phi_{3}}(\delta_{a3}-\phi_{a}\phi_{3})=0
\end{eqnarray}
\end{center}
which we re-write in terms of the acceleration of the field $\phi_a$:
\begin{equation}
\label{eom} \partial_{tt}{\phi}_{a}=K_{ab}^{-1} {\cal
  F}_b\left(\ph,\partial_{t}\ph,\partial_{i}\ph\right)
\end{equation}
with
\begin{equation}
K_{ab}=(1+2\theta_{S}\partial_{i}\ph\cdot\partial_{i}\ph)\delta_{ab}
-2\theta_{S}\partial_{i}\phi_a\partial_{i}\phi_b.
\end{equation}
We find that the inverse matrix of K exists in an explicit, but rather
messy form. The equation of motion is a second order PDE.

One drawback of the model is that the potential term is free for us to
choose. The most common choices are: $V=(1+\phi_3)^4$ (the holomorphic model chosen so that the 1-skyrmion solution is
$W=\lambda z$ (discussed in \cite{pz:93})),
$V=(1+\phi_3)$ (a 1-vacuum potential (studied  in
\cite{psz:95a}))
,$V=(1-\phi_3)(1+\phi_3)$ (a 2-vaccua potential (studied in
\cite{baby_pots})). Except for the first potential, no minimal-energy solutions are
known explicitly.
The baby Skyrme model is a non-integrable system and explicit solutions to
its resulting differential equations are nearly impossible to find. Numerical
methods are the only way forward.

\section{Numerical Techniques}

We are interested in two kinds of studies: the minimal-energy solution
in a given topological sector and the time-evolution of
configurations. We have used numerical techniques based on solving the
corresponding Euler-Lagrange equation. A randomisation  technique
e.g. Simulated Annealing might be more effective in finding the global
minimal energy solution: an initial field is randomly perturbed and
changes are selected via the Metropolis algorithm which accepts
changes to lower energy and allows  changes to higher energies according to a
certain probability. (see paper by Hale,  Schwindt and Weidig
\cite{sa})

We need (\ref{eom}) for the time-evolution
and relaxation of an initial configuration and use (\ref{profile}) to find
static hedgehog solutions. These DEs are re-written as sets of two first
order DEs. We discretise DEs by restricting our function to values at lattice
points and by reducing the derivatives to finite differences.\footnote{we use
the 9-point laplacian}(as explained in \cite{pz:98}, see also \cite{num}). We
take the time step to be half  the lattice spacing: $\delta
t=\frac{1}{2}\delta x$. We use fixed boundary conditions i.e.\ we set the
derivatives to zero at the boundary. We check our numerical results via
quantities conserved in the continuum limit and by changing lattice spacing
and number of points. Moreover, we compare them with theoretical predictions.

\paragraph{Minimal Energy solutions of the hedgehog Ansatz}

The static hedgehog solutions of (\ref{profile}) are found by the shooting
method using the 4th order Runge-Kutta integration and the boundary
conditions imposed by the finite energy condition. Alternatively, one can use a
relaxation technique like the Gauss-Seidel over-relaxation (\cite{num})
applied to an initial configuration with the same boundary conditions.  

\paragraph{Time-evolution and Non radially symmetric Minimal Energy Solutions}

The time-evolution of an initial configuration is determined
by the equation of motion (\ref{eom}). We are using the 4th order Runge-Kutta
method to evolve the initial set-up and correction techniques to keep the
errors small. We relax i.e.\ take out kinetic energy by using a damping (or
friction) term.
We construct $n$-skyrmions by relaxing an initial set-up of n 1-skyrmions
with relative phase shift of $\frac{\pi}{n}$. Using the dipole picture
developed by Piette et al. \cite{psz:95a}, two baby 1-skyrmions
attract each other for a non-zero value of the relative phase; phase shift of
$\frac{\pi}{2}$ for maximal attraction. A circular set-up is crucial as they
maximally attract each other and 1-skyrmions do not form several states that
repel each other.  We have run simulations for non-circular set-ups,
but either the 1-skyrmions take longer to fuse together or they fuse into
many bound-states and repel each other. We run our simulations on grids with
$200^2$ or $300^2$ lattice points and the lattice was $\delta x=0.1 \mbox{ or
} 0.05$. However, for large topological charge, we need larger grids and the
relaxation takes a long time. The corresponding hedgehog solution as an
initial set-up usually works well and is faster, but biased due to its large
symmetry group.

\paragraph{Initial set-up:} 
The initial field configuration is a linear superposition of static solutions
 with or without initial velocity; typically we use a circular set-up of n
 1-skyrmions with a $\frac{\pi}{n}$ phase-shift between each other (for
 maximal attraction).  The superposition is justified, because the profile
 function  decays exponentially. The superposition is done in the complex
 field formalism i.e.\ where $W$ is the stereographic projection of the $\ph$
 field of $S^2$ (see \cite{pz:93}). We use the profile function of a static
 solution (typically of topological charge one) to obtain  
\begin{equation}
  W=\tan\left(\frac{f(r)}{2}\right)e^{-in\theta}.
\end{equation}
This equation holds in the rest frame of a static skyrmion solution centred
around its origin and $\frac{dW}{dt}=0$. We may introduce moving solutions by
switching to a different frame of reference. This can be done by performing a
Lorentz boost on the rest frame of a given $W$, because $W$ is a Lorentz
scalar. It is  important that the different skyrmions are
not too close to each other.  Finally, the different $W$ fields are added together and  the complex field is re-written in
terms of the field $\ph$ and its derivative. 

\paragraph{Correction techniques on $S^2$:} 
The integration method introduces small errors which eventually add up.   In
terms of the $S^2$ constraint, this corresponds to the field leaving the
two-sphere. Hence, we need to project the field back onto the sphere.  The
simplest and sufficient projections are
$\phi_{a}\longrightarrow \frac{\phi_{a}}{\sqrt{\ph\cdot\ph}}$
and
$
\partial_{t}\phi_{a}\longrightarrow\partial_{t}\phi_{a}-
\frac{\partial_{t}\ph\cdot\ph}{\ph\cdot\ph}\phi_{a}$.
Of course, the space derivatives may also be corrected. See \cite{pz:98} for
further discussions.

\paragraph{Relaxation technique:} 
A damping term in the equation of motion will gradually take the kinetic
energy out of the system. The equation (\ref{eom}) changes to
\begin{equation}
\partial_{tt}{\phi}_{a}=K_{ab}^{-1} {\cal
  F}_b\left(\ph,\partial_{t}\ph,\partial_{i}\ph\right)
  -\gamma\partial_{t}\phi_a
\end{equation}
where $\gamma$ is the damping coefficient. We set $\gamma$ to 0.1, but most
values will do as long as they are not too large. Another approach would be
to absorb the outwards traveling kinetic energy waves in the boundary region.

\section{Numerical Studies}

We present three numerical studies of the baby Skyrme models.
\begin{itemize}

\item{\bf Check for Stability} 

It is important to check whether a solution of the Hedgehog Ansatz is
a global  minimum or not. We have taken the exact solution of the first
potential and perturbed it by a scale transformation.  The initial
configuration has twice the size of the static solution.  The skyrmion
`shakes' off its extra energy and radiates out kinetic energy
waves. The stable skyrmion remains; the stability of the 1-skyrmion is
fine and it probably is the minimum energy solution.

\item{\bf Scattering of baby skyrmions} 

First, the holomorphic baby skyrmions repel each other and only scatter if you collide them above a critical speed. Skyrmions of the 1-vacuum and 2-vacua potential have attractive channel and scatter in the same way. 
Figure~\ref{new_baby_scat_2} shows how the two 1-skyrmions
attract each other, form a bound-state, scatter away at 90 degrees, get
slowed down by their mutual attraction, attract each other again and so on.
This oscillating but stable bound-state is an excited state of the 2-skyrmion
solution. Taking out the kinetic energy, the bound-state relaxes to the
2-skyrmion; a ring.
\begin{figure}
\begin{center}
\caption{Contour plot of energy density: Scattering of 2 skyrmions}
\label{new_baby_scat_2}
\includegraphics[angle=0,scale=1,width=0.45\textwidth]{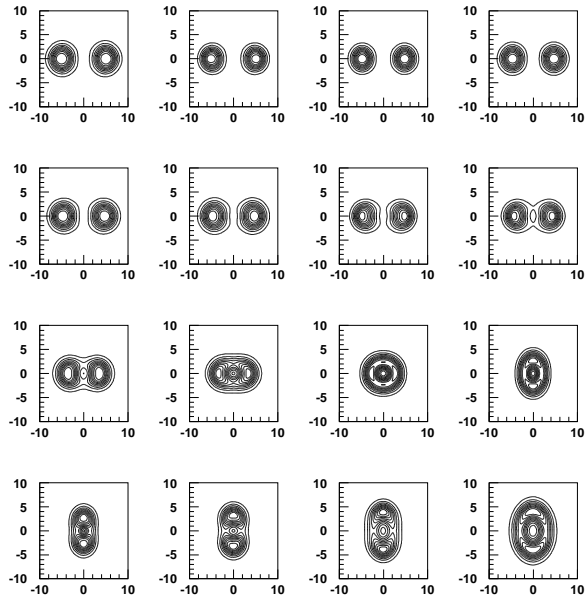}
\end{center}
\end{figure}

\item{\bf Multi-skyrmion solutions} 

The holomorphic model does not have
any  stable  n-skyrmions where $n$ is greater than one.
The common 1-vacuum potential possesses rather beautiful,
non-radially symmetric multi-skyrmions. They look like crystal with 
2-skyrmions as building blocks. (for example, see a 7-skyrmion in figure 
\ref{multi_7}.)
\begin{figure}
\begin{center}
\caption{Energy density of a 7-skyrmion}
\label{multi_7}
\includegraphics[angle=0,scale=1,width=0.4\textwidth]{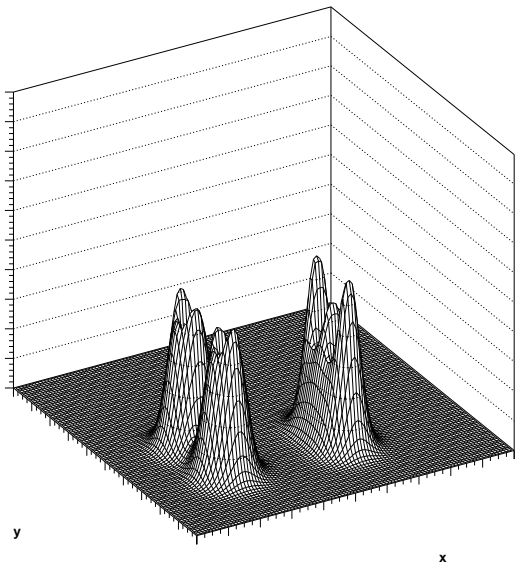}
\end{center}
\end{figure}
The 2-vacua potential gives rise to a remarkably different structure
for multi-skyrmions. In fact,  we show that the energy density of all
$n$-skyrmions turns out to be radially symmetric configurations,
namely rings of larger and larger radii. Figure \ref{new_baby_merge}
shows the relaxation to a `vibrating string' configuration (further relaxation leads to a perfect ring).
\begin{figure}
\begin{center}
\caption{Contour Plot of Energy Density: Formation of a 5-skyrmion}
\label{new_baby_merge}
\includegraphics[angle=0,scale=1,width=0.4\textwidth]{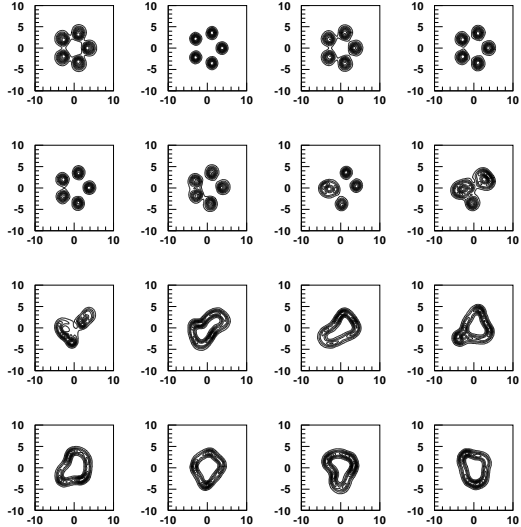}
\end{center}
\end{figure}
 Clearly, the choice of the potential term has a major impact on the
formation of multi-skyrmions and their shape. However, it is not clear
what controls the structure. (see our paper \cite{baby_pots})
\end{itemize}

\section{Conclusion}

We have studied classical aspects of the non-integrable baby Skyrme
models.  No exact solutions are known in general and we could only
achieve our study  by using numerical methods. We have shown that the
choice of potential term dictates the structure of the
multi-skyrmions. However, it is not clear to us what properties of
the potential controls the structure.  Our numerical approach is symptomatic 
for the increased use of computers in extracting valuable information from
theories untrackable by mathematical tools. `Numerical experiments'
are becoming a well established area between  `real' experiments and
pure theory.  Further, the author is in no doubt that solitons will
play an increasingly  important  role in fundamental
physics. Monopoles, instantons and  D-branes are other examples of
solitonic objects.  Physicists have two ways of thinking about
physical systems; in terms of waves or particles. Solitons combine
both fundamental concepts and are waves with particle behaviour. This
new concept is exciting and very appealing to theorists in all domains
-- a new tool to use with the help of computers.

First of all,  I want to express my gratitude to the directors of the
school Professors t'Hooft, Veneziano and Zichichi and Prof. Gourdin
for organising  an excellent and inspiring summer school. I also
congratulate Prof. t'Hooft on  his well deserved award of the Nobel
Price in Physics 1999. Further, I would like to thank the `Ettore
Majorana' Foundation and  Centre for Scientific Culture for the award
of the Sakurai Scholarship which paid the school fees and the
department of Mathematical Sciences at the University of Durham which
paid my travel expenses.  I also would like to thank my PhD supervisor
Wojtek Zakrzewski and  Bernard Piette for useful discussions and James
Stirling who selected me for  this summer school.

\bibliographystyle{plain} 
\bibliography{references}

\end{document}